\documentclass[aps,pra,showpacs,amsmath,amssymb,preprint]{revtex4} 

\usepackage{graphicx}
\usepackage{dcolumn}
\usepackage{bm}

\newcommand{ \micro }{ \mu }
\DeclareMathOperator{\Tr}{Tr}
\newcommand{\mean}[1]{{\ensuremath{\big<#1\big>}}}
\newcommand{\abs}[1]{\ensuremath{\left|#1\right|}}
\newcommand\relphantom[1]{\mathrel{\phantom{#1}}}

\begin{document}

\preprint{\today}
\title{Quantum kinetic theory model of a continuous atom laser}
\author{G.~R.\ Dennis$^1$, Matthew J. Davis$^2$, J. J. Hope$^1$}
\email{joseph.hope@anu.edu.au}
\affiliation{$^1$Australian National University, Department of Quantum Science, \\
ARC Centre of Excellence for Quantum-Atom Optics, ACT 0200, Australia\\
$^2$The University of Queensland, School of Mathematics and Physics, \\
ARC Centre of Excellence for Quantum-Atom Optics, QLD 4072, Australia.}

\begin{abstract}
We investigate the feasible limits for realising a continuously evaporated atom laser with high-temperature sources.  A plausible scheme for realising a truly continuous atom laser is to outcouple atoms from a partially condensed Bose gas, whilst continuously reloading the system with non-condensed thermal atoms and performing evaporative cooling.  Here we use quantum kinetic theory to model this system and estimate feasible limits for the operation of such a scheme.
For sufficiently high temperatures, the figure of merit for the source is shown to be the phase-space flux.  The dominant process limiting the usage of sources with low phase-space flux is the three-body loss of the condensed gas.  We conclude that certain double-magneto-optical trap (MOT) sources may produce substantial mean condensate numbers through continuous evaporation, and provide an atom laser source with a narrow linewidth and reasonable flux.\end{abstract}

\maketitle

\section{Introduction}

The experimental realisation of Bose-Einstein condensation (BEC) in dilute alkali gases \cite{BECreview} has opened up the study of the atom laser: a coherent beam of atoms analogous to the optical laser.  Like optical lasers, it is hoped that these atomic sources will demonstrate mode selectivity and high spectral flux \cite{Wiseman97}.  The simplest method of producing a spatially coherent atomic beam is to couple atoms out of a trapped BEC \cite{AtomLaserList}.  Rapid out-coupling of a BEC forms a coherent atomic beam with a spread in momentum as large as the trapped BEC, whereas coupling the atoms out more slowly reduces the output linewidth at the expense of reducing the overall flux \cite{MattiasPair}.  These atom lasers are equivalent to Q-switched optical lasers, which do not exhibit gain-narrowing.  Continuously pumping the atom laser to produce a stable output provides the obvious benefit of higher flux, but may also improve stability and linewidth of the output beam.  In a gain-narrowed optical laser, a higher pumping rate both increases the total flux and reduces the linewidth of the output, producing a dramatically increased spectral flux \cite{GN}.  An atom laser with gain-narrowing must have a saturable, Bose-enhanced pumping mechanism that operates simultaneously with the damping \cite{Wiseman97}.  This paper examines the limits on thermal sources that can be used to produce a continuous atom laser through the process of continuous evaporation.

The two essential steps towards providing a continuous pumping mechanism for an atom laser are: (1) the provision of atoms from an external source to the atomic trap, and (2) a process that causes at least some of those atoms to make an irreversible, stimulated transition into the BEC.  Continuous delivery of ultracold atoms has been demonstrated in a number of experiments \cite{Chikkatur:2002qa,Lahaye:2004,Greiner:2007,Muller:2007}, and is an important component of thermal atomic interferometry. 
Sequential reloading of a target BEC was achieved using optical tweezers~\cite{Chikkatur:2002qa}, where a series of source condensates were joined by manipulating the trapping potentials, and excitations were subsequently removed by further evaporative cooling. This milestone experiment maintained the condensate fraction, and therefore the potential flux of a potential atom laser.  However, an atom laser produced from such an experiment would not possess the desired narrow linewidth, as merging two coherent sources by manipulating the potential is not a Bose-enhanced process.  This means that the phase of the lasing mode would diffuse at a rate proportional to the replenishment rate.  In this paper it is shown that under certain conditions, a similar experiment using an ultra-cold \emph{thermal} source would be able to pump the target BEC and maintain a significant BEC population using a phase-preserving Bose-enhanced scattering process.

Atom-stimulated transitions into the condensate can be made irreversible by coupling to a reservoir. There are two possible reservoirs: the empty modes of the electromagnetic field accessible via a transition from an excited atomic state (as used in optical cooling), or the empty modes of the atomic field (accessible via evaporation).   Optical cooling is an obvious candidate for a continuous process, but it has not yet achieved BEC.  This is because condensates are very sensitive to the resonant light that is emitted by the spontaneous emission step in the cooling, as the recoil energy of a single photon is significantly larger than the energy per particle.  Several early atom laser models were based on optical cooling from a thermal source \cite{ALoptical}, but they have not been experimentally realised.  The only optical cooling method that has led to an increase in BEC number required a pre-condensed source in the same trap \cite{RobinsCts}.  

The last stage of cooling for production of all current BECs from thermal sources has been mediated by atom-atom interactions.  This has the advantage that it can be performed without the presence of resonant light, but the obvious disadvantage that it relies on the system approaching thermal equilibrium, and will therefore be reversed by the addition of atoms above the condensation temperature.  However, if the atoms can be supplied at the same time as the system undergoes a forced evaporation process, then even if the source atoms are above the condensation temperature, it is possible under some conditions to produce a net gain for the condensate mode.  It is therefore possible to make evaporative cooling operate in a continuous fashion.  The focus of this investigation is to determine exactly which existing thermal sources might be feasibly used for such an experiment.

The quantum kinetic theory of dilute gas BEC has been successfully used to model condensate growth \cite{QKTlist,Bijlsma:2000}.   We extend this model to include loss due to three-body inelastic scattering, which turns out to be a key determinant of the required phase-space flux required in the thermal source.  In Sec.~\ref{KineticTheory:SchemeModel}, we shall describe the schematic of our proposed pumping scheme, and the details of the quantum kinetic theory model.  The results of the model are examined in Sec.~\ref{KineticTheory:Results}.

\section{Scheme and kinetic theory model}
\label{KineticTheory:SchemeModel}

\begin{figure}
    \centering
        \includegraphics[width=10cm]{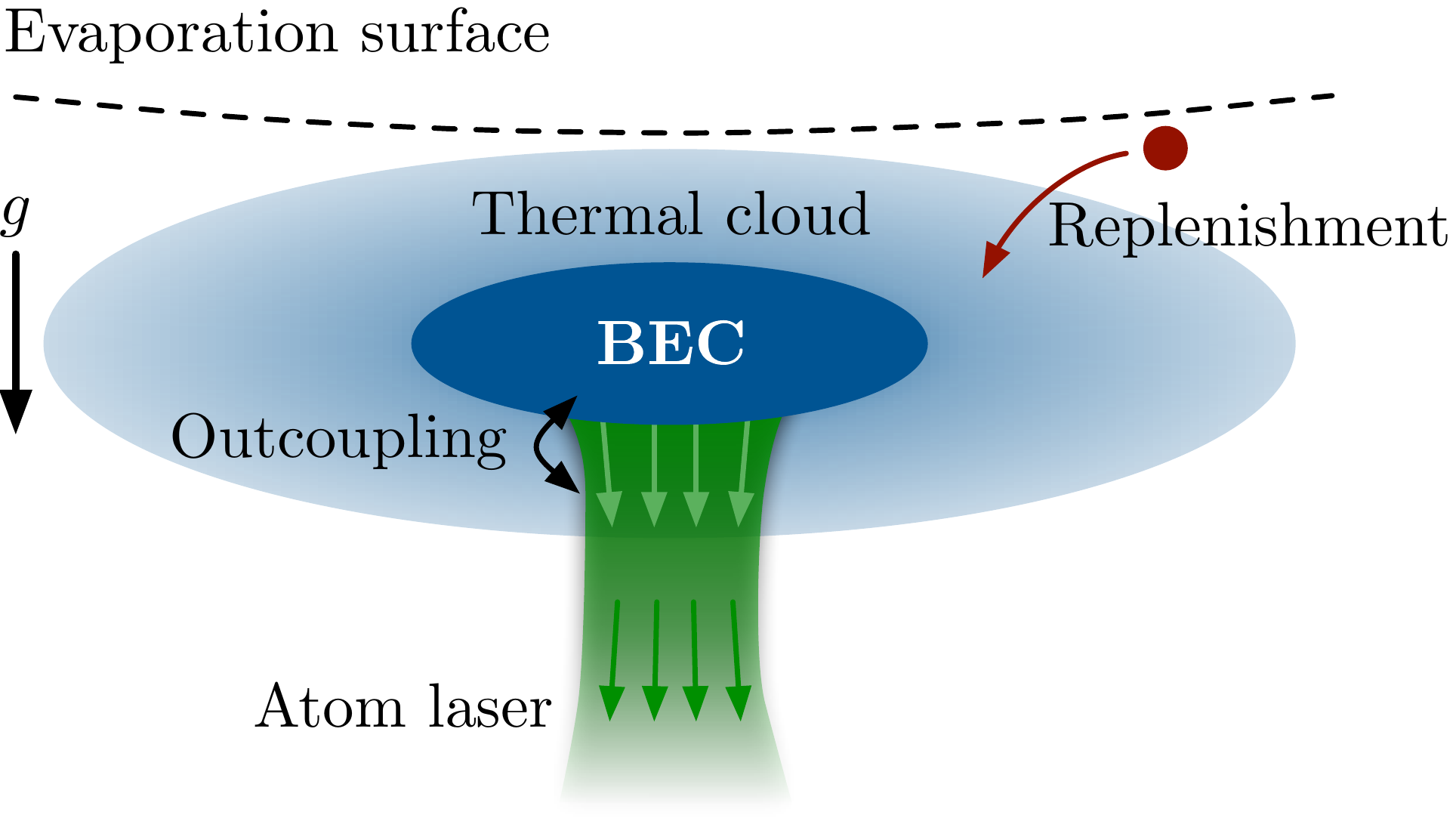}
    \caption{Schematic of the experimental setup.}
    \label{KineticTheory:QKTScheme}
\end{figure}

The proposed scheme for a pumped atom laser is illustrated in Fig.~\ref{KineticTheory:QKTScheme}.  It is very similar to the methods used to evaporate a thermal cloud to condensation in a magnetic trap and produce a (quasi-continuous) atom laser.  The gain process for the condensate is the same Bose-enhanced scattering between thermal atoms and the condensate that drives condensate growth when evaporating to produce BEC~\citep{Davis:2000vn,Bijlsma:2000,Gardiner:1997kx}.  This process becomes irreversible when one of the scattered atoms has enough energy to cross the evaporation surface and be removed from the thermal cloud.  The loss of atoms from the thermal cloud is balanced by a replenishment process that couples the thermal cloud to a source of atoms at finite temperature.  

The atom laser beam itself is produced by large momentum-transfer Raman outcoupling from the condensate.  Raman outcoupling improves the spatial properties of the beam \cite{ANUramanVsRF}, and allows minimal outcoupling from the thermal cloud when the two lasers are focussed to intersect only in the immediate vicinity of the condensate.

A non-equilibrium steady state will be reached when the rate of atom loss from the condensate due to outcoupling balances the rate of atoms gained due to growth collisions within the thermal cloud.  If the evaporative surface is tuned so that atoms of energy $\varepsilon_\text{cut}$ and higher are rapidly and continually removed from the trap, then all collisions that result in an atom with energy greater than $\varepsilon_\text{cut}$ will become irreversible. As $\varepsilon_\text{cut}$ is lowered, a larger fraction of the scattering processes that leave atoms in the condensate mode will become irreversible. This suggests that there must be some value of $\varepsilon_\text{cut}$ for which the condensate experiences net gain. What is not clear is whether the net gain can proceed efficiently, i.e. on a timescale much shorter than other losses from the condensate.  Lowering $\varepsilon_\text{cut}$ also reduces the total number of thermal atoms present in the steady state. In the limit that $\varepsilon_\text{cut}$ reaches the condensate energy, there will be no background gas at all, and the condensate cannot experience net gain.  We therefore expect that for a given set of parameters, there will be an optimal value for $\varepsilon_\text{cut}$ that maximises the net gain.  This value can be determined by a model based on quantum kinetic theory (QKT)~\cite{QKTlist}.

Our model is an extension of the kinetic models described in~\cite{Bijlsma:2000, Davis:2000vn}, which was successfully used to study condensate growth in an experiment in which a cloud of thermal atoms just above condensation temperature were shock-cooled through the BEC transition~\citep{Kohl:2002,Hugbart:2007}.  After cooling, the atoms were left to equilibrate, with condensate formation being driven by the same collisional processes that would drive condensate growth in the proposed pumped atom laser experiment described in the previous section.  To fully describe this proposed experiment, the kinetic model must be modified to include the effects of the replenishment and outcoupling processes, as well as the loss due to three-body recombination.  

Three-body recombination is the dominant loss process in typical BEC experiments~\cite{ThreeBody}, and it provides the main constraint on the size of the lasing mode.  
In fact, if the outcoupling from the condensate were the only non-negligible loss process from the trap, the largest condensate would be formed in the absence of evaporation.  Furthermore, the condensate number would be independent of the temperature of the replenishment source, depending only on the flux of atoms delivered to the system and the outcoupling rate from the condensate. This unphysical result is due to fact that in the absence of a density-dependent loss process, simply increasing the density is a feasible method of approaching degeneracy.   Density-dependent loss puts a limit on the physical density, and thus results in the expected trade-off between temperature and flux density into the model.

The starting point of the kinetic model is a separate treatment of the thermal and condensed components of the system.  The condensed component is assumed to be a quantum fluid obeying a Gross-Pitaevskii-type equation, however we make a further approximation and assume that the condensate is sufficiently occupied that it has a Thomas-Fermi profile \cite[Chapter 6]{PethickSmith}.  The condensate dynamics are then fully described by the number of condensed atoms $N_0(t)$.

The thermal cloud is described by the Hartree-Fock approximation~\citep[Chapter 8]{PethickSmith}, which assumes it is comprised of particle-like excitations moving in the effective potential of the harmonic trap plus condensate mean field.  To reduce the dimensionality of the full phase-space distribution function for the thermal cloud $f(\bm{r},\bm{p}, t)$, it is assumed that the system is ergodic, i.e.  that all points in the phase space having the same energy have equal population \cite{Luiten:1996}.  Under this approximation, the thermal cloud is then described by its energy distribution function $g(\varepsilon, t)$ and the density of states $\rho(\varepsilon, t)$.  The assumption of ergodicity has been shown in the past to give good agreement with experiment when asymmetric spatial or momentum dynamics are not significant~\citep{Bijlsma:2000,Davis:2000vn}.  Note that the time-dependence of the density of states $\rho(\varepsilon, t)$ comes from the contribution of the condensate mean field to the effective potential experienced by the thermal atoms.

As the model presented here is very similar to that presented in~\citep{Bijlsma:2000} with some additional terms, a derivation of the common terms is omitted. As a summary, the derivation proceeds by taking a semiclassical Boltzmann equation for the phase-space distribution function of the thermal cloud $f(\bm{r},\bm{p}, t)$ including collisional terms and using the ergodic approximation to obtain an equation of motion for the energy distribution function $g(\varepsilon, t)$. This equation is self-consistently matched with a Gross-Pitaevskii equation for the condensate before making the Thomas-Fermi approximation to obtain an equation of motion for the number of condensed atoms $N_0(t)$. An example application of this method to derive the appropriate terms for three-body loss is given in section \ref{MethodsAppendix:QKT3BodyLoss} of the appendix. Further, a detailed discussion of this theory is given in the review article~\citep{Proukakis:2008}.

\begin{figure}
    \centering
    \includegraphics[width=\columnwidth]{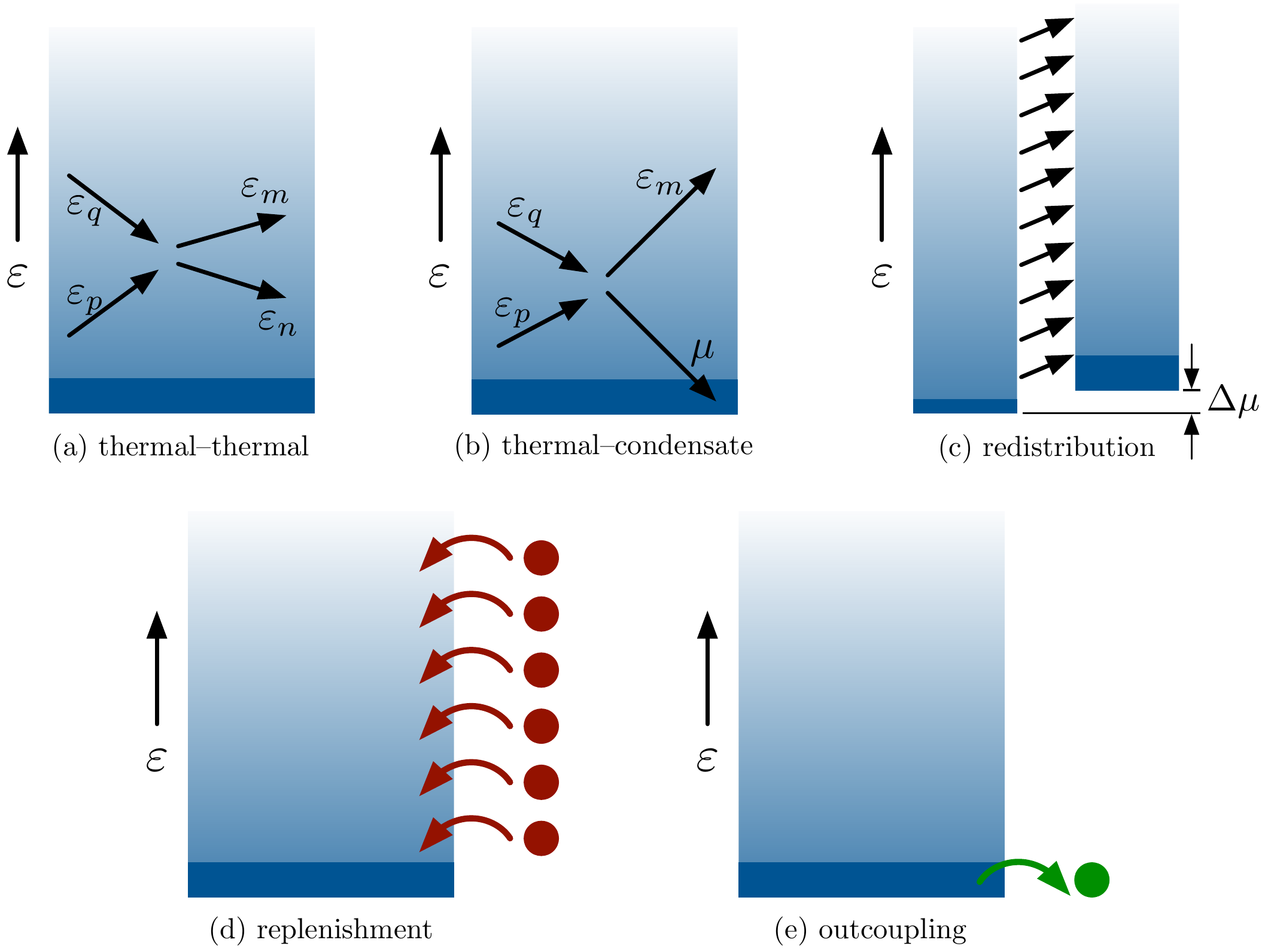}
    \caption{Schematic of processes involved in the evolution of the kinetic model described by \eqref{KineticTheory:EvolutionEquations}.  The upper shaded rectangle in each subfigure represents the energy distribution function $g(\varepsilon, t)$ of the thermal cloud, and the bottom dark blue rectangle represents the condensate with occupancy $N_0(t)$ and chemical potential $\varepsilon = \mu(t)$. Figures (a) and (b) represent collisional processes involving two thermal atoms and one thermal and one condensate atom respectively. Figure (c) represents the change in the energy distribution function $g(\varepsilon, t)$ if the condensate occupation (and hence chemical potential) increases, raising the energies of every energy level. Figure (d) represents the replenishment of the thermal cloud from an atomic reservoir, and Figure (e) represents outcoupling from the condensate mode to produce the atom laser.}
    \label{KineticTheory:ProcessDiagrams}
\end{figure}

Separating the contributions of the different processes involved, the equations of motion for the model for a collision-driven pumped atom laser considered here are
\begin{align}
    \frac{d N_0}{d t} =\begin{split}
        &\relphantom{+}\left. \frac{d N_0}{d t}\right|_\text{thermal--condensate} \\
        &+\left. \frac{d N_0}{d t}\right|_\text{3-body loss} \\
        &+\left. \frac{d N_0}{d t}\right|_\text{outcoupling}
    \end{split},
    & \frac{\partial (\rho g)}{\partial t} = \begin{split}
        &\relphantom{+}\left. \frac{\partial (\rho g)}{\partial t}\right|_\text{thermal--thermal} \\
        &+\left. \frac{\partial (\rho g)}{\partial t}\right|_\text{thermal--condensate} \\
        &+\left. \frac{\partial (\rho g)}{\partial t}\right|_\text{3-body loss} \\
        &+\left. \frac{\partial (\rho g)}{\partial t}\right|_\text{replenishment} \\
        &+\left. \frac{\partial (\rho g)}{\partial t}\right|_\text{redistribution}
    \end{split},
    \label{KineticTheory:EvolutionEquations}
\end{align}
where the subscripts `thermal--thermal' and `thermal--condensate' denote Bose-enhanced collisional processes between atoms in the corresponding states [Figs.~\ref{KineticTheory:ProcessDiagrams}(a) and (b), respectively], the subscript `3-body loss' indicates the contribution due to three-body recombination, the subscript `replenishment' indicates the contribution due to the replenishment of the thermal cloud [Fig.~\ref{KineticTheory:ProcessDiagrams}(d)], the subscript `outcoupling' indicates the contribution due to outcoupling from the condensate to form the atom laser [Fig.~\ref{KineticTheory:ProcessDiagrams}(e)], and the subscript `redistribution' indicates the contribution due to the redistribution of population in energy space due to the changes of the energies of the occupied levels as the mean-field of the condensate changes [Fig.~\ref{KineticTheory:ProcessDiagrams}(c)]. It is assumed that atoms with energy greater than the evaporative energy cut-off $\varepsilon_\text{cut}$ are removed from the system sufficiently quickly such that  $g(\varepsilon > \varepsilon_\text{cut}) = 0$.

The forms of the `thermal--thermal', `thermal--condensate' and `redistribution' terms in \eqref{KineticTheory:EvolutionEquations} are given in the appendix, and derivations are given in~\citep{Bijlsma:2000}.

The outcoupling process from the condensate is modelled as a simple linear loss process with corresponding rate constant $\gamma$,
\begin{align}
    \left.\frac{d N_0}{d t}\right|_\text{outcoupling} &= - \gamma N_0.
    \label{KineticTheory:OutcouplingProcess}
\end{align}
Modelling the outcoupling in this way neglects any outcoupling from thermal modes.  This is a reasonable approximation if focused Raman lasers are used for the outcoupling, which only intersect in the immediate vicinity of the condensate.  It also requires that the dominant losses from the thermal cloud are evaporation, coupling to the condensate and three-body loss. 

The thermal cloud is modelled as being continuously replenished from a source that provides a constant flux $\Phi$ of atoms at a temperature $T$.  To avoid tying the model to any particular replenishment mechanism, we assume a best-case scenario in which each energy level $\varepsilon$ in the source is coupled directly to the level in the thermal cloud with the same energy above the condensate chemical potential $\mu(t)$, i.e. the lowest energy level of the source ($\varepsilon=0$) is coupled directly to the lowest energy level in the trap ($\varepsilon = \mu(t)$).  This simple model gives the form of the contribution due to replenishment as
\begin{align}
    \left. \frac{\partial \big(\rho(\varepsilon, t) g(\varepsilon, t)\big)}{\partial t} \right|_\text{replenishment} &= \Gamma \rho_0(\varepsilon - \mu(t)) g_T(\varepsilon - \mu(t)),
    \label{KineticTheory:ReplenishmentProcess}
\end{align}
where $\rho_0(\varepsilon)$ is the density of states in the absence of a condensate, $g_T(\varepsilon)$ is the Bose-Einstein  distribution at temperature $T$, and $\Gamma$ is a rate constant such that
\begin{align}
    \Gamma \int_0^\infty \rho_0(\varepsilon) g_T(\varepsilon)\, d\varepsilon = \Phi,
    \label{KineticTheory:GammaPhiRelation}
\end{align}
where $\Phi$ is the flux of atoms from the source \emph{before} evaporation.  The derivation of the contributions to \eqref{KineticTheory:EvolutionEquations} due to three body loss are given in the appendix.

We summarise here the approximations made in obtaining the kinetic model \eqref{KineticTheory:EvolutionEquations}:
\begin{enumerate}
    \item The energy scale of the thermal cloud is large enough that all excitations are particle-like and not collective excitations such as phonons. Phonon-like excitations are only important for particle energies $\varepsilon \lesssim 2\mu(t)$~\citep[\S 8.3.1]{PethickSmith}. Hence, we require the that the energy scale for the thermal cloud $\varepsilon_\text{cut}$ be much larger than $\mu(t)$.
    \item The phase-space distribution of the thermal cloud is ergodic and hence is purely a function of energy. This assumption is true at equilibrium, however it needs some justification when used in non-equilibrium scenarios. In this case, asymmetric behaviour of the condensate is not expected in either position or momentum space.
    \item The condensate density is sufficiently large that it is well-described by a Thomas-Fermi profile. This approximation is justified as it is only the large-condensate limit that is of interest as a large condensate will be necessary for the production of a high-flux atom laser in this scheme. In making this approximation, the effects of both the normal and anomalous densities of the thermal cloud on the condensate have also been neglected.
    \item Evaporation occurs on a time-scale faster than collisions. This is the usual requirement during evaporation to condensation, and so should be satisfied in the proposed experiment.
\end{enumerate}

\section{Results}
\label{KineticTheory:Results}

For a given trap geometry, the model is fully defined by the flux of replenishment atoms $\Phi$, the temperature $T$ of those atoms, the energy of the evaporative cut $\varepsilon_\text{cut}$, and the outcoupling rate from the condensate $\gamma$. In this section the results of the kinetic model for some `typical' parameter values are presented, and the dependence of the model on each of the parameters is examined.

Our numerical simulations are based on a trap and conditions similar to that of~\citep{Kohl:2002}, who precooled a cloud of $^{87}$Rb atoms to an initial temperature slightly greater than the critical temperature before performing evaporative cooling to study condensate growth.  The trap in the experiment was axially-symmetric with radial and axial trapping frequencies of  $\omega_r = 2 \pi \times 110$ Hz and $\omega_z = 2\pi \times 14$ Hz respectively.

To solve the kinetic model numerically, Eq.~\eqref{KineticTheory:EvolutionEquations} is discretised along the energy dimension and the resulting coupled differential equations are solved with an adaptive fourth-fifth Runge-Kutta~\citep{NumericalRecipes} method. Our results are mainly concerned with the steady-state of the kinetic model, which we define as being reached when the condensate number has changed by less than $0.1\%$ or $1$ atom in $100$ ms.  The initial state for the simulation is chosen to be a truncated Bose-Einstein distribution containing (before truncation) $N_\text{initial} = 4.2\times 10^6$ atoms at the same temperature as the replenishment reservoir.  This state is chosen as a representation of the steady-state of the system prior to evaporation.  In the trap considered, the critical temperature for $4.2\times 10^6$ atoms is $T_c = 400$ nK.

\subsection{Typical results and parameter studies}
\label{SecKineticTheory:ParameterStudies}

\begin{figure}
    \centering
    \includegraphics[width=\columnwidth]{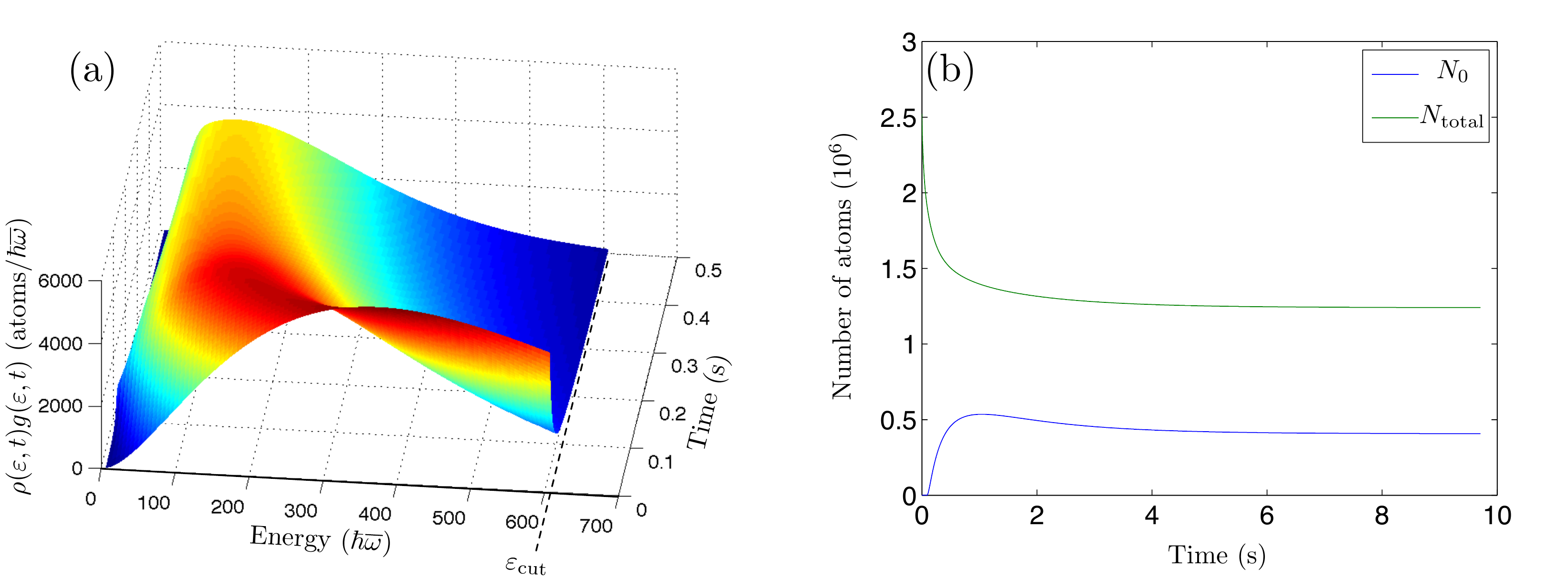}
    \caption{Results of the kinetic model for $\Phi = 8.4 \times 10^5$ atoms/s, $T=540$ nK, $\varepsilon_\text{cut} = 3 k_B T \approx 610  \hbar \overline{\omega}$, and $\gamma = 0.3$ s\textsuperscript{-1}. Figure (a) highlights the dynamics of the occupation of the thermal energy levels for $t < 0.5$ s, while (b) illustrates the equilibration of the total and condensed atom numbers over $\sim 10$ s. The energy distribution at $t=0$ s is a truncated Bose-Einstein distribution containing (before truncation) $N=4.2\times 10^6$~atoms at $T=540$ nK.
    }
    \label{KineticTheory:EnergyDistributionFunctionEvolution}
\end{figure}

As a depiction of the `typical' time-dependence of the results obtained from the kinetic theory model \eqref{KineticTheory:EvolutionEquations}, we consider the case of pumping the system continuously with a source such that the initial number $N = 4.2\times 10^6$ is transferred to the system once every 5 seconds giving a flux of $\Phi = 8.4 \times 10^5$~atoms/s.  The temperature of the replenishment source is chosen to be $T=540$ nK, 60\% above the condensation temperature of the system before evaporation.  For the remaining model parameters, we choose the evaporative cut-off to be $\varepsilon_\text{cut} = 3 k_B T$, and the outcoupling rate from the condensate to be $\gamma = 0.3$ s\textsuperscript{-1}.  

Figure \ref{KineticTheory:EnergyDistributionFunctionEvolution} illustrates the results of the simulation of this system.  Fig.~\ref{KineticTheory:EnergyDistributionFunctionEvolution}(a) shows the energy distribution of the thermal cloud cooling from the initial truncated Bose-Einstein distribution to a distribution with a lower average energy per particle.  Figure~\ref{KineticTheory:EnergyDistributionFunctionEvolution}(b) demonstrates that, despite pumping the system with an atomic reservoir above critical temperature, it is possible to reach a steady-state in which the condensate is macroscopically occupied.  In this example, the steady-state condensate fraction is 33\%.

\begin{figure}
    \centering
    \includegraphics[width= \columnwidth]{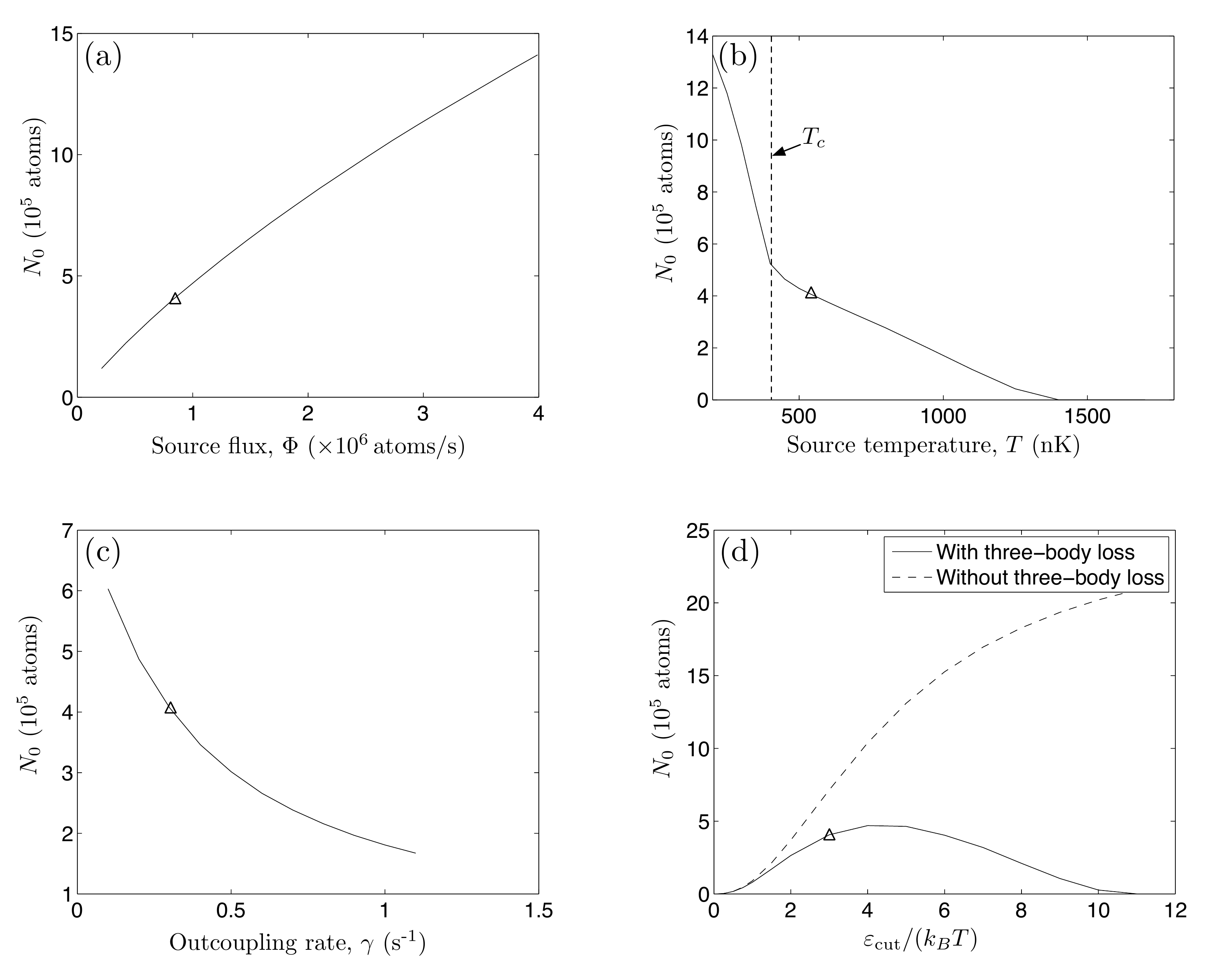}
    \caption{The dependence of the steady state condensate number $N_0$ on the parameters of the quantum kinetic model \eqref{KineticTheory:EvolutionEquations}. The steady state condensate number has a monotonic dependence on the replenishment flux $\Phi$ (a), the temperature T of the replenishment source (b) and the outcoupling rate $\gamma$ (c). For a given choice of the remaining parameters of the model there is an optimum $\varepsilon_\text{cut}$ (d) for which the equilibrium condensate number is a maximum. For each parameter being varied, the remaining parameters are chosen to be the same as for the results depicted in Fig.~\ref{KineticTheory:EnergyDistributionFunctionEvolution}. The triangle in each plot marks the point that corresponds to the precise conditions of Fig.~\ref{KineticTheory:EnergyDistributionFunctionEvolution} in steady state.}
    \label{KineticTheory:ParameterStudies}
\end{figure}

The non-equilibrium dynamics of the system are not the subject of investigation here, instead our interest is in the steady state itself, and in determining the feasibility of creating a pumped atom laser driven by a non-condensed atomic source.  As a first step towards this investigation we consider the dependence of the equilibrium condensate number on the parameters of the system: $\Phi$, $T$, $\varepsilon_\text{cut}$, and $\gamma$.  The results of this parameter study are presented in Fig.~\ref{KineticTheory:ParameterStudies}, where we use the initial conditions of Fig.~\ref{KineticTheory:EnergyDistributionFunctionEvolution}.

In general the parameter dependences depicted in Fig.~\ref{KineticTheory:ParameterStudies} are straightforward; adjusting each parameter causes a monotonic change in the equilibrium condensate number.  Increasing the flux of atoms to the system increases the equilibrium condensate number [Fig.~\ref{KineticTheory:ParameterStudies}(a)], while increasing the temperature of the replenishment source or increasing the outcoupling rate reduces the equilibrium condensate number [Fig.~\ref{KineticTheory:ParameterStudies}(b) and (c) respectively].  The only behaviour that is not straightforward is displayed by Fig.~\ref{KineticTheory:ParameterStudies}(d) in which the dependence on the evaporative cut-off $\varepsilon_\text{cut}$ is illustrated.  For large $\varepsilon_\text{cut}$, few atoms will be lost due to evaporation and the system will reach equilibrium when the flux of atoms into the system is balanced by three-body losses and outcoupling from the condensate.  As $\varepsilon_\text{cut}$ is reduced, more atoms are lost due to evaporation and the mean energy per particle reduces causing the condensate size to increase.  As $\varepsilon_\text{cut}$ continues to decrease, an increasing fraction of the replenishment atoms have an energy greater than $\varepsilon_\text{cut}$ causing a lower effective atomic flux to be delivered to the system, hence reducing the steady state size of any condensate formed.  These two competing effects are the origin of the existence of an optimum steady state condensate number as a function of $\varepsilon_\text{cut}$ in Fig.~\ref{KineticTheory:ParameterStudies}(d).

As discussed earlier, in the absence of three-body loss the equilibrium condensate number would continue to increase as $\varepsilon_\text{cut}$ is increased, which would lead to the unphysical conclusion that evaporating \emph{reduces} the equilibrium condensate number.  This is demonstrated by the dashed line in Fig.~\ref{KineticTheory:ParameterStudies}(d) which asymptotes towards $N_0=\Phi/\gamma = 2.8\times 10^6$ atoms in the limit $\varepsilon_\text{cut}\rightarrow \infty$.  As observed in the remaining panels of Fig.~\ref{KineticTheory:ParameterStudies} (in which the effects of three-body loss have been included) three-body loss does not give rise to optimum values for the corresponding parameters of the model as it is only changes to $\varepsilon_\text{cut}$ that affect the evaporative and three-body losses in contrary fashions.  An increase in the replenishment flux will increase both evaporative and three-body losses.  Similarly, changes to the temperature of the replenishment source or the outcoupling rate either increase or decrease both of the evaporative and three-body losses.

At this point it is clear that to have the largest equilibrium condensate number one should use a replenishment source with the highest possible flux and the lowest possible temperature.  However any such recommendation would be disingenuous.  It is an experimental reality that these parameters are not entirely uncoupled; while a $300$ K oven might produce a significantly larger flux than a $50$ mK 2D-MOT, it is simply not realistic to create a $50$ mK atomic source with the same flux as the $300$ K oven.  We will now examine this trade-off between the temperature and flux in the context of experimentally-realisable sources.



\subsection{Behaviour in the high-temperature limit}
In the previous section we investigated the dependence of the equilibrium condensate number on the model parameters. The physical question that we wish to address with this model is: What are the limits of a thermal atom source such that is can realise a pumped, continuous atom laser?

Although it would be possible to create a pumped atom laser by combining condensates in a manner similar to the experiment by Chikkatur~\emph{et al.} \cite{Chikkatur:2002qa}, such an atom laser would have significantly reduced phase-stability unless the replenishment process were essentially continuous. However, to replenish a condensate by collisional interactions with a continuous source of condensed atoms, the replenishment source would itself need to have many of the desired properties of a pumped atom laser.  Instead, it would be preferable to use a source \emph{above} condensation temperature for replenishment.

We consider now the experimentally-relevant limit of replenishing the thermal cloud using a high-flux source of thermal atoms. For such sources two simplifications are possible. First, for temperatures greater than $T_c$ the Bose-Einstein energy distribution of the source $g_T(\varepsilon)$ is well approximated by the Boltzmann distribution $g_T(\varepsilon) \approx \zeta e^{-\beta \varepsilon}$ for some constant $\zeta$, and $\beta = \left(k_B T\right)^{-1}$. Secondly, for high temperature sources the optimum evaporation cut-off $\varepsilon_\text{cut}$ will be much smaller than the characteristic energy of the source $k_B T$, and hence $\varepsilon_\text{cut} \ll k_B T$.  From these simplifications it can be seen that the energy distribution below the evaporation cut-off is well described by the single parameter $\zeta$ as $g_T(\varepsilon \leq \varepsilon_\text{cut}) \approx \zeta$.

At this point we have simply rewritten the temperature dependence of the replenishment source in terms of the parameter $\zeta$.  However, as the energy distribution of the replenishment source only affects the kinetic model through Eq.~\eqref{KineticTheory:ReplenishmentProcess}, its influence on the system dynamics is only through the combined quantity $\kappa = \Gamma \zeta$. An approximate expression for $\kappa$ in the Boltzmann limit directly in terms of relevant experimental quantities can be obtained using the definition Eq.~\eqref{KineticTheory:GammaPhiRelation},
\begin{align}
    \Phi &= \Gamma \int_0^\infty \rho_0(\varepsilon) g_T(\varepsilon)\, d\varepsilon\\
    &= \Gamma \int_0^\infty \frac{\varepsilon^2}{2 (\hbar \overline{\omega})^3} \zeta e^{-\beta \varepsilon}\, d\varepsilon\\
    &= \Gamma \zeta \frac{1}{2 (\hbar \overline{\omega})^3} \int_0^\infty \varepsilon^2 e^{-\beta \varepsilon}\, d\varepsilon\\
    &= \left(\frac{k_B T}{\hbar \overline{\omega}}\right)^3 \Gamma \zeta\\
    \kappa &\equiv \Gamma \zeta = \Phi \left(\frac{\hbar \overline{\omega}}{k_B T}\right)^3 \label{KineticTheory:KappaDefinition}
\end{align}
where $\overline{\omega} = \left(\omega_x \omega_y \omega_z\right)^{\frac{1}{3}}$ is the geometric mean of the trapping frequencies, and $\displaystyle\rho_0(\varepsilon) = \frac{\varepsilon^2}{2 (\hbar \overline{\omega})^3}$ is the density of states in a harmonic trap in the absence of a condensate~\citep{PethickSmith}.

We identify $\kappa$ as the \emph{phase-space flux} of the source as it is directly related to the rate at which the phase-space density of the thermal source is delivered.  For a trap of $N$ thermal atoms at temperature $T$, the peak phase-space density $\varpi$ is~\citep[Chapter 2]{PethickSmith}
\begin{align}
    \varpi &= N \left(\frac{\hbar \overline{\omega}}{k_B T} \right)^3.
\end{align}
If these $N$ atoms are delivered over a time $\tau$ providing a flux $\Phi = N/\tau$ the peak phase-space flux is
\begin{align}
    \frac{\varpi}{\tau} &= \frac{N}{\tau}\left(\frac{\hbar \overline{\omega}}{k_B T}\right)^3 = \Phi \left(\frac{\hbar \overline{\omega}}{k_B T}\right)^3 \equiv \kappa.
\end{align}
The parameter $\zeta$ is therefore the phase-space density $\varpi$.

The phase-space flux $\kappa$ is a figure-of-merit for the thermal source.  It quantifies the qualitative behaviour already known: for the same atomic flux $\Phi$, a source with a lower temperature will result in a larger condensate [Fig.~\ref{KineticTheory:ParameterStudies}(b)]; and for the same temperature, a source with a higher atomic flux will also result in a larger condensate [Fig.~\ref{KineticTheory:ParameterStudies}(a)].  The phase-space flux also describes exactly how a trade-off between the flux and temperature of the replenishment source will affect the equilibrium condensate number.  If two sources with different fluxes and temperatures have the same value phase-space flux, then the equilibrium condensate number produced by the two sources will be the same (assuming the high-temperature limit applies to both sources).  Our interest is in determining what values of $\kappa$ are necessary to produce a pumped atom laser, and whether such values are achievable.

For the limit of high-temperature atomic sources, we have reduced the four variables $(\Phi, T, \varepsilon_\text{cut}, \gamma)$ required to define the model \eqref{KineticTheory:EvolutionEquations} down to three $(\kappa, \varepsilon_\text{cut}, \gamma)$. Of these three, our main interest is in the dependence of the system on the properties of the atomic source through $\kappa$. In contrast, the dependence of the equilibrium condensate number on the outcoupling rate $\gamma$ is simple (see Fig.~\ref{KineticTheory:ParameterStudies}(c)) and the results would not be expected to change qualitatively with $\gamma$. It is therefore appropriate to choose a representative value for the outcoupling rate (here $\gamma = 0.3$ s\textsuperscript{-1}) and focus on the remaining two quantities.

\begin{figure}
    \centering
    \includegraphics[width=7cm]{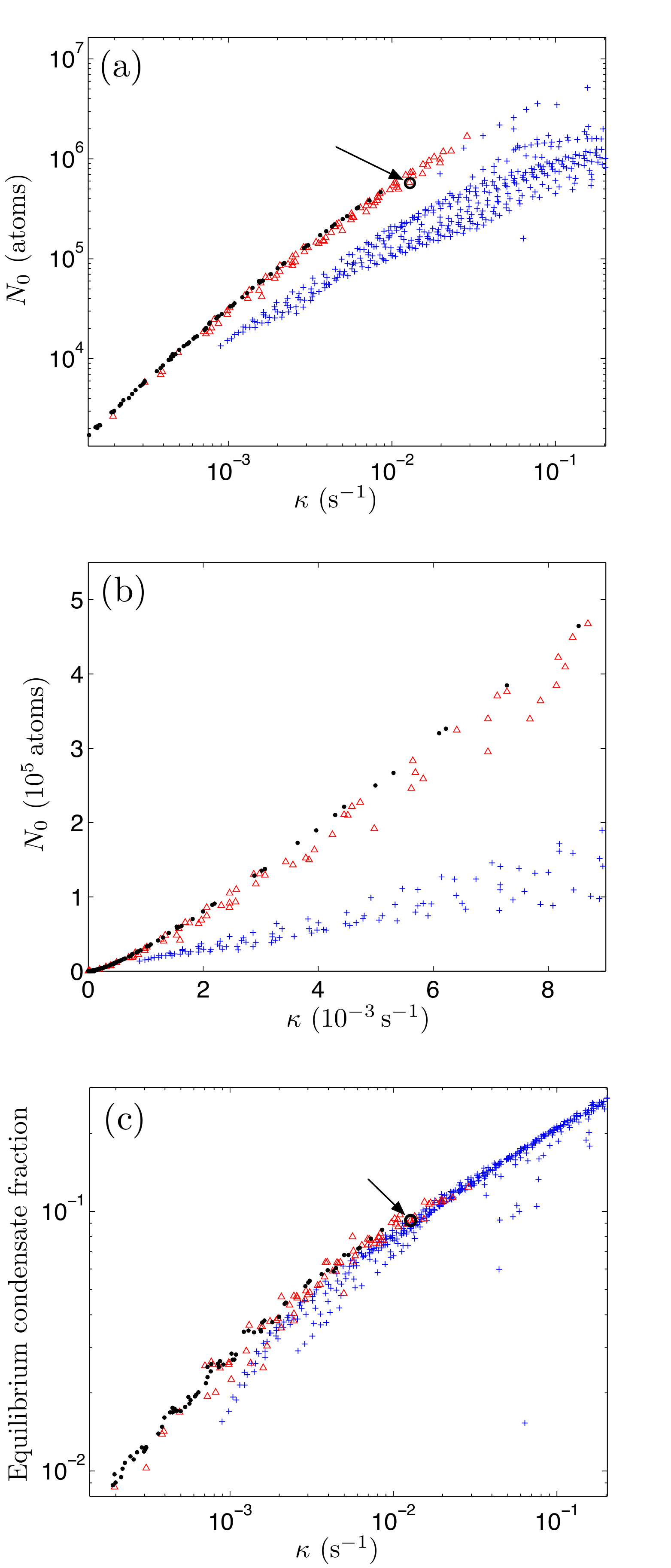}
    \caption{Steady-state condensate properties as a function of the phase-space flux $\kappa$ for the replenishment source. Figures (a) and (b) plot the steady-state condensate number on linear-linear and log-log scales. Figure (c) plots the equilibrium condensate fraction $N_0/N$.  The results of the parameter scan are divided into three groups (the black circles, red triangles and blue crosses) based on their proximity to the high-temperature limit, where $\kappa$ is expected to be the only figure of merit. The circled point with the arrow pointing to it corresponds to a simulation of the parameters for the last source in Table \ref{KineticTheory:ExperimentalSources}, which has $\kappa= 1.1\times 10^{-2}$ s\textsuperscript{-1} (see main text).}
    \label{KineticTheory:FigureOfMerit}
\end{figure}

As discussed in the previous section, there is an optimal choice for the evaporative cut-off $\varepsilon_\text{cut}$. Our interest here is in the best-case scenario: for a given thermal source, what is the largest condensate we can produce? To examine this question and to verify that $\kappa$ does fully describe the properties of the thermal source in the appropriate limit we have performed a parameter scan of the model \eqref{KineticTheory:EvolutionEquations} for a range of fluxes $1.3\times 10^5$ s$^{-1} < \Phi < 5\times 10^{10}$ s\textsuperscript{-1} and temperatures $200$ nK $< T < 600 \micro$K of the atomic source, for each combination determining the optimum evaporative cut $\varepsilon_\text{cut}$ to give the largest steady-state condensate number.  The results of this parameter scan are displayed in Fig.~\ref{KineticTheory:FigureOfMerit}.

The results illustrated in Fig.~\ref{KineticTheory:FigureOfMerit} are separated into three groups based on the ratio $\varepsilon_\text{cut}/(k_B T)$.  The first group marked by black circles have $\varepsilon_\text{cut}/(k_B T) < 0.1$, and are the results for which the high-temperature limit can be considered to be a good approximation (satisfying the requirement $\varepsilon_\text{cut} \ll k_B T$).  Here $\kappa$ completely determines the properties of the replenishment source.  For this group of results, any equilibrium property of the system should appear to be a single (not necessarily straight) line when plotted as a function of $\kappa$.  The results in Fig.~\ref{KineticTheory:FigureOfMerit} demonstrate that these results can be viewed as a single function of $\kappa$.  The second group marked by red triangles have $0.1 \leq \varepsilon_\text{cut}/(k_B T) < 0.5$, and can be considered to be the results for which the high-temperature limit is almost a good approximation.  These results are reasonably close to the results of the first group, however there is a greater deviation for a given value of $\kappa$ indicating that the results can be almost seen as purely a function of $\kappa$. All remaining results fall into the third group for which $\varepsilon_\text{cut}/(k_B T) \geq 0.5$.  It can be seen that these points correspond to a broad range of steady states, indicating that the replenishment source cannot be described by $\kappa$ alone.

The first two panels of Fig.~\ref{KineticTheory:FigureOfMerit} both display the equilibrium condensate number as a function of the phase-space flux $\kappa$.  Fig.~\ref{KineticTheory:FigureOfMerit}(a) uses a log-log scale to highlight the behaviour for small and large values of $\kappa$, while Fig.~\ref{KineticTheory:FigureOfMerit}(b) uses a linear-linear scale to demonstrate that the black circles lying on a single line in Fig.~\ref{KineticTheory:FigureOfMerit}(a) are not simply an artefact of plotting the results using a logarithmic scale.  Finally, Fig.~\ref{KineticTheory:FigureOfMerit}(c) displays the equilibrium condensate fraction as a function of $\kappa$.

Figures \ref{KineticTheory:FigureOfMerit}(a) and (b) demonstrate that it would be possible to produce atom lasers with respectable condensate numbers $N_0 \gtrsim 10^5$ (corresponding to atom laser fluxes of $\gtrsim 3\times 10^4$ atoms/s for the outcoupling rate $\gamma = 0.3$ s\textsuperscript{-1} chosen) by using replenishment sources that have a phase-space flux $\kappa \gtrsim 10^{-3}$ s\textsuperscript{-1}.  To determine if this is experimentally feasible, the properties of a range of experimental atomic sources are detailed in Table \ref{KineticTheory:ExperimentalSources} and the corresponding values of the phase-space flux $\kappa$ calculated.

\begin{table}
    \begin{minipage}{\textwidth}
        \renewcommand{\footnoterule}{}
        \centering
        \begin{tabular}{ p{3.8cm} c c c r}
        \toprule
         & Atomic flux   & Temperature  & Phase-space flux  &\\
        Atomic source &  $\Phi$  &  $T$ &  $\kappa$ &\\
        \toprule 
        BECs in dipole traps & $10^5 \text{ s\textsuperscript{-1}}$ ($55 \text{ mHz}$)\footnote{This source is pulsed, and the flux is the mean flux over one cycle with the repetition rate listed in parentheses.}\footnote{This repetition rate it too low for this source to be useful (see main text). It is listed for purposes of comparison only.} & $<1 { \micro \text{K}}$ & $>1.9\times 10^{-3} \text{ s\textsuperscript{-1}}$ &~\citep{Chikkatur:2002qa} \\
        2D\textsuperscript{+}-MOT & $9\times 10^{9}\text{ s\textsuperscript{-1}}$ & $38 \text{ mK}$\footnote{In keeping with the best-case scenario investigation being performed, this temperature assumes that the mean velocity of the atoms can be reduced to zero without affecting the distribution. This could be achieved, for example, by firing the source vertically below the main pumped atom laser experiment and taking the atoms from the mean turning point.}\footnote{The dominant contribution to this temperature is the spread in the longitudinal velocities of the atoms.} & $3 \times 10^{-12} \text{ s\textsuperscript{-1}}$ &~\citep{Dieckmann:1998} \\
        2D\textsuperscript{+}-MOT & $2\times 10^{10}\text{ s\textsuperscript{-1}}$ & $42\text{ mK}$\textsuperscript{\emph{cd}} & $5 \times 10^{-12}{\text{ s}\textsuperscript{-1}}$ &~\citep{Chaudhuri:2006} \\
        MM-MOT & $10^9{\text{ s}\textsuperscript{-1}}$ & $61{ \micro \text{K}}$\textsuperscript{\emph{c}} & $8 \times 10^{-5}\text{ s\textsuperscript{-1}}$ &~\citep{Cren:2002rt}\\
        LVIS & $5\times 10^{9}\text{ s\textsuperscript{-1}}$ & $25\text{ mK}$\textsuperscript{\emph{cd}} & $6 \times 10^{-12}\text{ s\textsuperscript{-1}}$ &~\citep{Lu:1996} \\
        Zeeman slower & $3.2\times 10^{12}\text{ s\textsuperscript{-1}}$ & $32\text{ mK}$\textsuperscript{\emph{cd}} & $2\times10^{-9}\text{ s\textsuperscript{-1}}$ &~\citep{Slowe:2005} \\
        Magnetic guide loaded from 3D-MOT & $7\times 10^{9}\text{ s\textsuperscript{-1}}$ & $400 \micro\text{K}$\textsuperscript{\emph{c}} & $2\times 10^{-6}\text{ s\textsuperscript{-1}}$ &~\citep{Lahaye:2004}\\
        3D-MOT loaded from Zeeman slower & $2\times 10^{10}\text{ s\textsuperscript{-1}}$ ($0.5\text{ Hz}$)\textsuperscript{\emph{a}} & $500 \micro\text{K}$ & $3\times 10^{-6}\text{ s\textsuperscript{-1}}$ &~\citep{Streed:2006}\\
        3D-MOT loaded from 2D\textsuperscript{+}-MOT & $3\times 10^8\text{ s\textsuperscript{-1}}$ ($3\text{ Hz}$)\textsuperscript{\emph{a}} & $8 \micro\text{K}$ & $1.1\times 10^{-2}\text{ s\textsuperscript{-1}}$ &~\citep{Muller:2007}\\
        \toprule 
        \end{tabular}
    \end{minipage}
    \caption{Relevant properties of selected experimental cold atomic sources. The phase-space flux $\kappa$ is evaluated from the atomic flux and temperature values listed using \eqref{KineticTheory:KappaDefinition}.}
    \label{KineticTheory:ExperimentalSources}
\end{table}

The first source listed in Table \ref{KineticTheory:ExperimentalSources} is the experiment of Ref.~\cite{Chikkatur:2002qa} that merged independently produced BECs in optical dipole traps that was discussed earlier.  While this experiment is certainly not in the high-temperature limit, it has been included for comparison purposes.  Of the remaining sources listed in Table \ref{KineticTheory:ExperimentalSources}, most are many orders of magnitude away from being useful potential sources for a pumped atom laser (cf.\ Fig.~\ref{KineticTheory:FigureOfMerit}).  The fluxes obtainable from these sources are insufficient to compensate for their higher temperatures as an increase of three-orders of magnitude in flux is necessary to compensate for an increase of a single order of magnitude in temperature (see \eqref{KineticTheory:KappaDefinition}).  Only the last atomic source satisfies the requirement $\kappa \gtrsim 10^{-3}$s\textsuperscript{-1}.  This experiment by M\"{u}ller \emph{et al.}~\cite{Muller:2007} is one of the sources in a dual atom interferometer designed for the precision measurement of accelerations and rotations~\citep{Muller:2009}.  A direct simulation has been performed for the parameters of this source, and the results are marked by a circle with an arrow pointing to it in Fig.~\ref{KineticTheory:FigureOfMerit}.  

The steady-state condensate number for the source of M\"{u}ller \emph{et al.}~\cite{Muller:2007} is $N_0 = 5 \times 10^5$ atoms, which would be a sufficiently large condensate to serve as a stable phase-reference for an atom laser, were it a pure BEC.  However the equilibrium condensate fraction for this source is only 10\% (see \ref{KineticTheory:FigureOfMerit}(c)).  Previous theoretical work investigating the transfer of statistics from a trapped (quasi-)condensate to an atom laser found that using high-momentum kick Raman outcoupling such as that proposed in the scheme presented here can filter some of these fluctuations causing the atom laser to have a larger coherence length than the condensate from which it was produced~\citep{Proukakis:2003}.  

It is not possible to investigate the transfer of statistics from the trapped component to the atom laser within the present model due to the simplifying assumption that it is only the condensate mode that is outcoupled to form the atom laser.  A more detailed three-dimensional model taking into account the full spatial dependence of the Raman outcoupling process would be necessary to fully determine the feasibility of using an atomic source such as that described by M\"{u}ller \emph{et al.}~\cite{Muller:2007} in the production of a truly continuous pumped atom laser.



\section{Conclusions and Outlook}

We have investigated the feasibility of producing a continuously pumped atom laser fed by evaporation and replenishment with a thermal source.  The method has been to investigate the best-case scenario in which the replenishment process introduces no heating to the trapped thermal component beyond that due to bringing the replenishing atoms into contact with the thermal cloud.  With these caveats in mind, the results are promising: using an existing experimental source~\citep{Muller:2007} it appears possible to produce steady-state condensates with large atom number ($\sim 5\times 10^5$ atoms) using the scheme presented in Fig.~\ref{KineticTheory:QKTScheme}.  Could the atomic flux of this source be increased by an order of magnitude, the condensate number produced by this scheme could be pushed to $5\times 10^6$ atoms.  

Ultimately it is not the size of the condensate that we are interested in, but the resulting flux of the atom laser and its coherence length.  The former goal will certainly be improved by larger equilibrium condensate sizes, however the atom laser itself will be useless unless its coherence length is sufficiently large compared to MOT thermal sources to offset the reduced atom flux.  It is by no means clear from the present work whether an atom laser produced using the atomic source of~ M\"{u}ller \emph{et al.}~\cite{Muller:2007} will have a coherence length that is larger than the size of the condensate from which it was outcoupled.  To investigate this point it will be necessary to not make the assumption of ergodicity and to include the full spatial dependence of the Raman outcoupler.  This could be achieved by making use of a finite temperature classical field theory~\citep{Blakie:2008a}.

One thing is certainly clear from the results presented here: the replenishment source for a collisionally-pumped atom laser must be close to degeneracy. It is simply not realistic to compensate for higher temperatures with a sufficiently increased flux.


\begin{acknowledgments}
This research was supported by the Australian Research Council Centre of Excellence for Quantum-Atom Optics (ACQAO). We acknowledge the use of CPU time at the National Computational Infrastructure National Facility.
\end{acknowledgments}

\appendix*
\section[Details of the Quantum Kinetic Theory model]{Details of the Quantum Kinetic Theory model}
\label{MethodsAppendix:KineticTheory}

One of the difficulties involved in solving the kinetic model is that the energy range that the problem is defined over changes in time.  The maximum energy is simply the energy of the evaporative cut-off $\varepsilon_\text{cut}$, while the minimum energy is the chemical potential of the condensate $\mu(t)$.  A discretisation of the energy dimension over the range $[0, \varepsilon_\text{cut}]$ will suffer from problems accurately representing the lower-end of the distribution where the minimum energy of the thermal atoms is varying.  An alternative is write the problem in terms of a shifted energy coordinate $\overline{\varepsilon} \equiv \varepsilon - \mu(t)$ so that the minimum energy of the system is now fixed \cite{Bijlsma:2000}.  Of course the same problem now exists at the upper end of the energy range where the maximum energy $\overline{\varepsilon}_\text{max} = \varepsilon_\text{cut} - \mu(t)$ is now time-dependent.  However, at equilibrium there will be significantly fewer thermal atoms at the evaporation cut-off than there will be near the condensate (this is illustrated in Fig.~\ref{KineticTheory:EnergyDistributionFunctionEvolution}(a)).  This choice will then result in smaller numerical errors than the alternative.

Written in terms of the shifted energy variable $\overline{\varepsilon}$, the contribution due to the replenishment is
\begin{align}
    \left. \frac{\partial\big(\overline{\rho}(\overline{\varepsilon}, t) \overline{g}(\overline{\varepsilon}, t)\big)}{\partial t}\right|_\text{replenishment} &= \Gamma \overline{\rho}_0(\overline{\varepsilon}) \overline{g}_T(\overline{\varepsilon}),
    \label{MethodsAppendix:KineticTheory:ReplenishmentProcess}
\end{align}
where a bar over a function is used to indicate that it is defined in terms of the shifted energy coordinate.  The original form of this term is given by \eqref{KineticTheory:ReplenishmentProcess}.

\subsection{Density of states}
\label{MethodsAppendix:QKTDensityOfStates}

Although the evolution equations for the kinetic model \eqref{KineticTheory:EvolutionEquations} are written in terms of the product of the density of states $\rho(\varepsilon, t)$ and the energy distribution function $g(\varepsilon, t)$, it will be necessary to separately determine the energy distribution function to evaluate the collisional contributions given in the following section. To extract the energy distribution function it is necessary to have an explicit expression for the density of states for the thermal cloud. This density of states is not simply the same as that for a harmonic trap as the thermal modes will experience a mean-field repulsion due to the condensate mode. The effective potential experienced by the thermal atoms is
\begin{align}
    V_\text{eff}(\bm{r}, t) &= V_\text{trap}(\bm{r}) + 2 g n_c(\bm{r}, t), \label{MethodsAppendix:QKTEffectivePotential}
\end{align}
where $V_\text{trap}(\bm{r})$ is the potential due to the magnetic trap, $g = 4\pi \hbar^2 a/m$, $a$ is the \emph{s}-wave scattering length and $n_c(\bm{r}, t)$ is the condensate density which was assumed to follow a Thomas-Fermi distribution.  Note that the `2' in the above expression is the full Hartree-Fock mean field experienced by the thermal atoms (see \citep[Chapter 8]{PethickSmith} for further details) which is twice the mean-field repulsion experienced by condensate atoms.  This is essentially due to the thermal atoms being distinguishable from the condensate atoms, while the condensate atoms are indistinguishable from one another.

The density of states in the presence of the effective potential \eqref{MethodsAppendix:QKTEffectivePotential} is given by
\begin{align}
    \rho(\varepsilon, t) &= \int \frac{d \bm{r}\, d \bm{p}}{(2\pi \hbar)^3} \,\delta\left(\varepsilon - V_\text{eff}(\bm{r}, t) - \bm{p}^2/2 m\right).
    \label{MethodsAppendix:DensityOfStatesDefinition}
\end{align}
The integrals are performed in~\citep{Bijlsma:2000} giving the following result in terms of the shifted energy coordinate (Eqs.~49 and 50 in~\citep{Bijlsma:2000})
\begin{align}
    \overline{\rho}(\overline{\varepsilon}, t) &= \frac{2}{\pi \hbar \overline{\omega}} \left[I_-(\overline{\varepsilon}) + I_+(\overline{\varepsilon})\right],
\end{align}
where the functions $I_\pm(\overline{\varepsilon})$ are
\begin{align}
    I_-(\overline{\varepsilon}) &= \left.\frac{u_-^3 x}{4} - \frac{a_- u_- x}{8} - \frac{a_-^2}{8}\ln(x + u_-)\right|_{x=\sqrt{\max\{0, -a_-\}}}^{x=\sqrt{2\mu/\hbar \overline{\omega}}}, \label{MethodsAppendix:QKTIMinus}\\
    I_+(\overline{\varepsilon}) &= \left.- \frac{u_+^3 x}{4} + \frac{a_+ u_+ x}{8} + \frac{a_+^2}{8} \arcsin\left(\frac{x}{\sqrt{a_+}}\right)\right|_{x=\sqrt{2\mu/\hbar\overline{\omega}}}^{x=\sqrt{a_+}}, \label{MethodsAppendix:QKTIPlus}
\end{align}
with $a_\pm = 2(\overline{\varepsilon}\pm \mu)/\hbar\overline{\omega}$, and $u_\pm = \sqrt{a_\pm \mp x^2}$. Note that there is a minor typo in Bijlsma \emph{et al.}~\cite{Bijlsma:2000}, the lower limit of $I_-(\overline{\varepsilon})$ is given as $x=\sqrt{\max\{0, a_-\}}$, while it should read $x=\sqrt{\max\{0, -a_-\}}$ as in \eqref{MethodsAppendix:QKTIMinus}.

\subsection{Collision and energy-redistribution in Quantum Kinetic Theory}
\label{MethodsAppendix:QKTOtherTerms}

A full derivation of the forms of the collision and energy-redistribution terms of the kinetic model is given in~\citep{Bijlsma:2000,Proukakis:2008}.

The contribution due to thermal--thermal collisions is given in Eq.~26 of~\citep{Bijlsma:2000} and has the form
\begin{align}
    \begin{split}
        \left. \frac{\partial\big(\overline{\rho}(\overline{\varepsilon}_1, t) \overline{g}(\overline{\varepsilon}_1, t)\big)}{\partial t}\right|_\text{thermal--thermal} &= \frac{m^3 g^2}{2 \pi^3 \hbar^7} \int d\overline{\varepsilon}_2 \int d\overline{\varepsilon}_3 \int d\overline{\varepsilon}_4 \,\overline{\rho}(\overline{\varepsilon}_\text{min}, t)\\
        &\phantom{=} \times\delta(\overline{\varepsilon}_1 + \overline{\varepsilon}_2 - \overline{\varepsilon}_3 - \overline{\varepsilon}_4) \\
        &\phantom{=} \times [ (1+\overline{g}_1) (1+\overline{g}_2) \overline{g}_3 \overline{g}_4 - \overline{g}_1 \overline{g}_2 (1+\overline{g}_3) (1+\overline{g}_4)],
    \end{split}
\end{align}
where $\overline{\varepsilon}_\text{min}$ is the minimum of the $\overline{\varepsilon}_i$, and $\overline{g}_i = \overline{g}(\overline{\varepsilon}_i, t)$. 

The contribution due to thermal--condensate collisions is given by Eq.~53 and Eq.~58--60 of~\citep{Bijlsma:2000} and has the form
\begin{align}
    \begin{split}
        \left. \frac{\partial\big(\overline{\rho}(\overline{\varepsilon}_1, t) \overline{g}(\overline{\varepsilon}_1, t)\big)}{\partial t}\right|_\text{thermal--condensate} &= \frac{m^3 g^2}{2 \pi^3 \hbar^7} \int d \overline{\varepsilon}_2 \int d\overline{\varepsilon}_3 \int d\overline{\varepsilon}_4 \,\delta(\overline{\varepsilon}_2 - \overline{\varepsilon}_3 - \overline{\varepsilon}_4)\\
        &\phantom{=}\times\left[ \delta(\overline{\varepsilon}_1 - \overline{\varepsilon}_2) - \delta(\overline{\varepsilon}_1 - \overline{\varepsilon}_3) - \delta(\overline{\varepsilon}_1- \overline{\varepsilon}_4)\right]\\
        &\phantom{=} \times \left[(1+\overline{g}_2)\overline{g}_3 \overline{g}_4 - \overline{g}_2 (1+\overline{g}_3)(1+\overline{g}_4) \right]\\
        &\phantom{=}\times  \int_{\overline{U}_\text{eff}(\bm{r}, t) \leq \overline{U}_-} d \bm{r}\, n_c(\bm{r}, t),
    \end{split}
    \label{MethodsAppendix:QKTRhoGThermalCondensateEvolution}
\end{align}
where $\displaystyle \overline{U}_- = \frac{2}{3}\left[(\overline{\varepsilon}_3 + \overline{\varepsilon}_4)-\sqrt{\overline{\varepsilon}_3^2 - \overline{\varepsilon}_3 \overline{\varepsilon}_4 + \overline{\varepsilon}_4^2}\right]$, and $\overline{U}_\text{eff}(\bm{r}, t) = U_\text{eff}(\bm{r}, t) - \mu(t)$.
The corresponding contribution to the evolution of the condensate number is simply
\begin{align}
    \left. \frac{d N_0}{d t}\right|_\text{thermal--condensate} &= - \int d\overline{\varepsilon} \,\left. \frac{\partial\big(\overline{\rho}(\overline{\varepsilon}, t) \overline{g}(\overline{\varepsilon}, t)\big)}{\partial t}\right|_\text{thermal--condensate}.
    \label{MethodsAppendix:QKTNThermalCondensateEvolution}
\end{align}

Finally, the contribution due to energy redistribution is (Eqs.~32 and 52 in~\citep{Bijlsma:2000})
\begin{align}
    \left. \frac{\partial\big(\overline{\rho}(\overline{\varepsilon}_1, t) \overline{g}(\overline{\varepsilon}_1, t)\big)}{\partial t}\right|_\text{redistribution} &= - \frac{\partial \big( \overline{\rho}_\text{w} \overline{g}\big)}{\partial \overline{\varepsilon}},
    \label{MethodsAppendix:QKTRedistributionEvolution}
\end{align}
where $\overline{\rho}_\text{w}$ is the weighted density of states
\begin{align}
    \overline{\rho}_\text{w}(\overline{\varepsilon}) &= \frac{2}{\pi \hbar \overline{\omega}} \left[ I_-(\overline{\varepsilon}) - I_+(\overline{\varepsilon})\right] \frac{d \mu}{dt},
\end{align}
where the functions $I_\pm(\overline{\varepsilon})$ are given in \eqref{MethodsAppendix:QKTIMinus} and \eqref{MethodsAppendix:QKTIPlus}.

\subsection{Three-body loss in Quantum Kinetic Theory}
\label{MethodsAppendix:QKT3BodyLoss}

The dominant density-dependent loss process in Bose-Einstein condensates is three-body loss~\citep{Burt:1997fk,Soding:1999}.  Three-body loss (or three-body recombination) is the process in which three atoms collide forming a bound dimer with the third necessary to ensure both energy and momentum conservation.  The binding energy is sufficient to give the products of a three-body recombination process sufficient kinetic energy to rapidly escape the trap.  Three-body loss is then well-described by the master equation term
\begin{align}
    \left. \frac{d \hat{\rho}}{d t}\right|_\text{3-body loss} &= \frac{1}{3}L_3 \int d \bm{x} \,\mathcal{D} \left[ \hat{\Psi}^3(\bm{x}) \right] \hat{\rho},
    \label{MethodsAppendix:ThreeBodyLossMasterEquationTerm}
\end{align}
where $\mathcal{D}[\hat{c}]\hat{\rho} = \hat{c}\hat{\rho} \hat{c}^\dagger - \frac{1}{2}(\hat{c}^\dagger \hat{c}\hat{\rho} + \hat{\rho} \hat{c}^\dagger \hat{c})$ is the usual decoherence superoperator, and $L_3 = 5.8\times 10^{-30}\text{cm\textsuperscript{6}s\textsuperscript{-1}}$~\citep{Burt:1997fk} is the three-body recombination loss rate constant.  This equation, first derived by~\citep{Jack:2002} has the familiar form of a decoherence superoperator with the state undergoing loss as the argument. 

The loss rate of atoms from the system due to three-body loss is readily obtained from \eqref{MethodsAppendix:ThreeBodyLossMasterEquationTerm} as
\begin{align}
    \left.\frac{d N}{dt} \right|_\text{3-body loss} &=  \Tr\left\{\int d \bm{r}\,\hat{\Psi}^\dagger(\bm{r})\hat{\Psi}(\bm{r}) \left.\frac{d \hat{\rho}}{dt}\right|_\text{3-body loss} \right\} = - L_3 \int d \bm{r}\, \mean{\hat{\Psi}^\dagger(\bm{r})^3 \hat{\Psi}(\bm{r})^3}.
    \label{MethodsAppendix:ThreeBodyLossNumberLossRate}
\end{align}

To separate the contributions to \eqref{MethodsAppendix:ThreeBodyLossNumberLossRate} due to the thermal and condensed components, we use a broken symmetry approach.  We write the annihilation operator $\hat{\Psi}$ in terms of its mean value $\Psi \equiv \mean{\hat{\Psi}}$ and the fluctuation operator $\delta \hat{\Psi} \equiv \hat{\Psi} - \Psi$ and substitute this into \eqref{MethodsAppendix:ThreeBodyLossNumberLossRate}.  The fluctuation operator defined here includes thermal fluctuations, which cannot be considered to be small.  Higher powers of $\delta\hat{\Psi}$ can therefore not be neglected.  However, thermal fluctuations have no well-defined phase relationship to one another or to the condensate.  Expectation values containing an unequal number of creation and annihilation deviation operators such as $\mean{\delta\hat{\Psi}\delta\hat{\Psi}}$ can therefore be assumed to be zero.

Performing the substitution described, \eqref{MethodsAppendix:ThreeBodyLossNumberLossRate} becomes
\begin{align}
    \begin{split}
        \left.\frac{d N}{dt} \right|_\text{3-body loss} &=  -L_3 \int d \bm{r}\, \Big\{[n_c(\bm{r})]^3 + 9 [n_c(\bm{r})]^2 \mean{\delta\hat{\Psi}^\dagger(\bm{r}) \delta\hat{\Psi}(\bm{r})}\\
        &\phantom{=-L_3 \int d \bm{r}\,\Big\{} + 9 n_c(\bm{r}) \mean{\delta\hat{\Psi}^\dagger(\bm{r})^2 \delta\hat{\Psi}(\bm{r})^2} + \mean{\delta\hat{\Psi}^\dagger(\bm{r})^3 \delta\hat{\Psi}(\bm{r})^3}\Big\},
    \end{split}
    \label{MethodsAppendix:ThreeBodyLossNumberLossRateInTermsOfDeviationOperators}
\end{align}
where $n_c(\bm{r}) = \abs{\Psi(\bm{r})}^2$ is the condensate density.

The non-condensate density is given by $n_T(\bm{r}) = \mean{\delta\hat{\Psi}^\dagger(\bm{r}) \delta\hat{\Psi}(\bm{r})}$.  As thermal states are Gaussian, the higher-order expectation values in the previous expression may be simplified by the application of Wick's theorem~\citep{Wick:1950} giving
\begin{align}
    \mean{\delta\hat{\Psi}^\dagger(\bm{r})^2 \delta\hat{\Psi}(\bm{r})^2} &= 2 [n_T(\bm{r})]^2, \\
    \mean{\delta\hat{\Psi}^\dagger(\bm{r})^3 \delta\hat{\Psi}(\bm{r})^3} &= 6 [n_T(\bm{r})]^3.
\end{align}
Substituting these expressions back into \eqref{MethodsAppendix:ThreeBodyLossNumberLossRateInTermsOfDeviationOperators} yields
\begin{align}
    \left.\frac{d N}{dt} \right|_\text{3-body loss} &=  -L_3 \int d \bm{r}\, [n_c(\bm{r})]^3 + 9 [n_c(\bm{r})]^2 n_T(\bm{r}) + 18 n_c(\bm{r}) [n_T(\bm{r})]^2 + 6 [n_T(\bm{r})]^3.
    \label{MethodsAppendix:ThreeBodyLossNumberLossRateInTermsOfDensities}
\end{align}

The evaluation of this loss rate requires the evaluation of the condensate and thermal densities.  The condensate density $n_c(\bm{r})$ is fully determined by the condensate occupation $N_0(t)$ within the Thomas-Fermi approximation that has already been made elsewhere in the derivation of the kinetic model.  The first term of \eqref{MethodsAppendix:ThreeBodyLossNumberLossRateInTermsOfDensities} only involves the condensate density and may be evaluated analytically
\begin{align}
    \frac{d N_0}{dt} &= - L_3 \frac{15^{4/5}}{168 \pi^2} \left(\frac{m \overline{\omega}}{\hbar \sqrt{a}} \right)^{12/5} N_0^{9/5}.
\end{align}
The remaining terms of \eqref{MethodsAppendix:ThreeBodyLossNumberLossRateInTermsOfDensities} require an expression for the thermal density $n_T(\bm{r})$, which can be obtained from the energy distribution function $g(\varepsilon)$ and the density of states $\rho(\varepsilon)$.

The total number of thermal atoms $N_T$ can be written as
\begin{align}
    N_T &= \int d\varepsilon\, \rho(\varepsilon) g(\varepsilon),
    \label{MethodsAppendix:NThermalDefinition}
\end{align}
where the density of states is defined by \eqref{MethodsAppendix:DensityOfStatesDefinition}.  Substituting this into \eqref{MethodsAppendix:NThermalDefinition} and rearranging the order of integrals gives
\begin{align}
    N_T &= \int d \bm{r}\, \int d\varepsilon\, \rho(\varepsilon, \bm{r}) g(\varepsilon),
    \label{MethodsAppendix:NThermalInTermsOfRhoER}
\end{align}
where we have defined
\begin{align}
    \rho(\varepsilon, \bm{r}) &= \int d \bm{p}\, \delta\left(\varepsilon - V_\text{eff}(\bm{r}, t) - \bm{p}^2/2m\right) = \frac{m^{3/2}}{\sqrt{2} \pi^2 \hbar^3} \sqrt{\varepsilon - V_\text{eff}(\bm{r})}.
\end{align}
The thermal density can be identified from \eqref{MethodsAppendix:NThermalInTermsOfRhoER}
\begin{align}
    n_T(\bm{r}) &= \int d\varepsilon\, \rho(\varepsilon, \bm{r}) g(\varepsilon).
    \label{MethodsAppendix:nThermalInTermsOfRhoER}
\end{align}

The remaining terms of \eqref{MethodsAppendix:ThreeBodyLossNumberLossRateInTermsOfDensities} can now be expressed in terms of the energy distribution function $g(\varepsilon)$ and the density of states $\rho(\varepsilon)$ by substituting \eqref{MethodsAppendix:nThermalInTermsOfRhoER} for one of the factors of $n_T(\bm{r})$ in each term
\begin{align}
    -L_3 \int d \bm{r}\, 9 [n_c(\bm{r})]^2 n_T(\bm{r}) &= - L_3 \int d\varepsilon\, g(\varepsilon) \int d \bm{r}\, 9 \rho(\varepsilon, \bm{r}) [n_c(\bm{r})]^2, \label{MethodsAppendix:Condensate2Thermal1}\\
    -L_3 \int d \bm{r}\, 18 n_c(\bm{r}) [n_T(\bm{r})]^2 &= - L_3 \int d\varepsilon\, g(\varepsilon) \int d \bm{r}\, 18 \rho(\varepsilon, \bm{r}) n_c(\bm{r}) n_T(\bm{r}), \label{MethodsAppendix:Condensate1Thermal2}\\
    -L_3 \int d \bm{r}\, 6 [n_T(\bm{r})]^3 &= - L_3 \int d\varepsilon\, g(\varepsilon) \int d \bm{r}\, 6 \rho(\varepsilon, \bm{r}) [n_T(\bm{r})]^2. \label{MethodsAppendix:Condensate0Thermal3}
\end{align}
From these expressions the rate of loss of atoms of energy $\varepsilon$ from the distribution can be identified
\begin{align}
    \left.\frac{\partial\big(\rho(\varepsilon) g(\varepsilon)\big)}{\partial t}\right|_\text{3-body loss} &= -L_3 \int d \bm{r}\, \rho(\varepsilon, \bm{r}) g(\varepsilon) \left\{ 3 [n_c(\bm{r})]^2 + 12 n_c(\bm{r}) n_T(\bm{r}) + 6 [n_T(\bm{r})]^2 \right\},
    \label{MethodsAppendix:QKT3BodyLossDistributionEvolution}
\end{align}
where the contributions due to the terms involving only one or two thermal atoms have been multiplied by $1/3$ and $2/3$ respectively to share appropriately the total loss.  The corresponding term for the condensate number evolution is
\begin{align}
    \begin{split}
        \left. \frac{d N_0}{dt} \right|_\text{3-body loss} &= - L_3 \frac{15^{4/5}}{168 \pi^2} \left(\frac{m \overline{\omega}}{\hbar \sqrt{a}} \right)^{12/5} N_0^{9/5}\\
        &\relphantom{=} - L_3 \int d\varepsilon \int d \bm{r}\, \rho(\varepsilon, \bm{r}) g(\varepsilon) \left\{6 [n_c(\bm{r})]^2 + 6 n_c(\bm{r}) n_T(\bm{r}) \right\},
    \end{split}
    \label{MethodsAppendix:QKT3BodyLossCondensateEvolution}
\end{align}
where the contributions due to the terms involving only one or two condensate atoms have been multiplied by $1/3$ and $2/3$ respectively.


\begin{thebibliography}{99}

\bibitem{BECreview}F. Dalfovo \textit{et al.}, \rmp {\bf 71}, 463 (1999); A. J. Leggett, \rmp {\bf 73}, 307 (2001); M. H. Anderson {\em et al.}, Science {\bf 269}, 198 (1995); K. B. Davis {\em et al.}, \prl {\bf 75},  3969(1995); C. C. Bradley, C. A. Sackett, J. J. Tollett, and R. G. Hulet, \prl {\bf 75}, 1687 (1995).
\bibitem{Wiseman97}H. M. Wiseman, \pra {\bf 56}, 2068 (1997).
\bibitem{AtomLaserList}M.-O. Mewes, M. R. Andrews, D. M. Kurn, D. S. Durfee, C.~G. Townsend, and W. Ketterle, \prl {\bf 78}, 582 (1997); J.-F. Riou, W. Guerin, Y. Le Coq, M. Fauquembergue, V.~Josse, P. Bouyer, and A. Aspect, \prl {\bf 96}, 070404 (2006); N. P. Robins, C. Figl, S. A. Haine, A. K. Morrison, M. Jeppesen, J. J. Hope, and J. D. Close, \prl {\bf 96}, 140403 (2006); A. …ttl, S. Ritter, M. Kšhl, and T. Esslinger, \prl {\bf 95},  090404 (2005).
\bibitem{MattiasPair}M. Johnsson, S. Haine, J.J. Hope, N.P. Robins, C. Figl, M. Jeppesen, J. DuguŽ, and J.D. Close, \pra {\bf 75}, 043618 (2007); M. Johnsson and, J.J. Hope, \pra {\bf 75}, 043619 (2007).
\bibitem{GN}L. Mandel and E. Wolf, \textit{Optical Coherence and Quantum Optics}, (Cambridge University Press, Cambridge, 1995).
\bibitem{Chikkatur:2002qa}A.P. Chikkatur, Y. Shin, A.E. Leanhardt, D. Kielpinski, E. Tsikata, T.L. Gustavson, D.E. Pritchard, and W. Ketterle, Science {\bf 296}, 2193 (2002).
\bibitem{Muller:2007}T.~M\"{u}ller, T.\ Wendrich, M.\ Gilowski, C.\ Jentsch, E.~M.\ Rasel, and W.\ Ertmer, \pra {\bf 76}, 063611 (2007).
\bibitem{Lahaye:2004}T.\ Lahaye, J.~M.\ Vogels, K.~J.\ G\"{u}nter, Z.\ Wang, J.\ Dalibard, and D.\ Gu\'{e}ry-Odelin, \prl {\bf 93}, 093003 (2004).
\bibitem{Greiner:2007} A.\ Greiner, J.\ Sebastian, P.\ Rehme, A.\ Aghajani-Talesh, A.\ Griesmaier, and T.\ Pfau, J.\ Phys.\ B \textbf{40}, F77 (2007).
\bibitem{ALoptical}H.~M. Wiseman and M.~J. Collett, Phys. Lett. A {\bf 202},  246  (1995); S.Bhongale and M.Holland, \pra {\bf 62}, 043604 (2000); L. Santos {\em et al.}, Phys. Rev. A {\bf 63}, 063408 (2001).
\bibitem{RobinsCts}N.P. Robins, C. Figl, M. Jeppesen, G.R. Dennis, and J.D. Close, Nat. Phys. {\bf 4}, 731 (2008).
\bibitem{Bijlsma:2000}M.J. Bijlsma, E. Zaremba, and H.T.C. Stoof, \pra {\bf 62}, 063609 (2000).
\bibitem{QKTlist} C.~W.\ Gardiner and P.\ Zoller, \pra \textbf{55}, 2902 (1997); D.\ Jaksch, C.~W.\ Gardiner, and P.\ Zoller, \pra \textbf{56}, 575 (1997); C.~W.\ Gardiner and P.\ Zoller, \pra \textbf{58}, 536 (1998); D.\ Jaksch, C.~W.\ Gardiner, K.~M.\ Gheri, and P.\ Zoller, \pra \textbf{58}, 1450 (1998); C.~W.\ Gardiner and P.\ Zoller, \pra \textbf{61}, 033601 (2000); M.~D.\ Lee and C.~W.\ Gardiner, \pra \textbf{62}, 033606 (2000).
\bibitem{Davis:2000vn}M.J. Davis, C.W. Gardiner, and R.J. Ballagh, \pra {\bf 62}, 063608 (2000).
\bibitem{Gardiner:1997kx}C.W. Gardiner, \pra {\bf 56}, 1414 (1997).
\bibitem{ANUramanVsRF}N.P. Robins, C. Figl, S.A. Haine, A.K. Morrison, M. Jeppesen, J.J. Hope, and J.D. Close, \prl {\bf 96}, 140403 (2006).
\bibitem{Miesner:1998}H.-J. Miesner, D.~M. Stamper-Kurn, M.~R. Andrews, D.~S. Durfee, S.~Inouye, and W.~Ketterle, Science {\bf 279}, 1005 (1998)
\bibitem{ThreeBody}E.A. Burt, R.W. Ghrist, C.J. Myatt, M.J. Holland, E.A. Cornell, and C.E. Wieman, Phys. Rev. Lett. 79, 337 (1997); J. S\"{o}ding, D. Gu{\'e}ry-Odelin, P. Desbiolles, F. Chevy, H. Inamori, and J. Dalibard, Applied Physics B: Lasers and Optics {\bf 69}, 257 (1999).
\bibitem{PethickSmith}C.~J. Pethick and H.~Smith, {\em Bose-Einstein Condensation in Dilute Gases} (Cambridge University Press, 2002).
\bibitem{Luiten:1996}O.~J. Luiten, M.~W. Reynolds, and J.~T.~M. Walraven, \pra {\bf 53}, 381 (1996).
\bibitem{Proukakis:2008}N.~P. Proukakis and B.~Jackson, J.\ Phys.\ B {\bf 41}, 203002 (2008).
\bibitem{Kohl:2002}M.~K\"{o}hl, M.~J.\ Davis, C.~W.\ Gardiner, T.~W.\ H\"{a}nsch, and T.\ Esslinger, \prl {\bf 88}, 080402 (2002).
\bibitem{Hugbart:2007}M.~Hugbart, J.~A. Retter, A.~Var˜n, P.~Bouyer, A.~Aspect and M.~J.\ Davis, \pra {\bf 75}, 011602(R) (2007).
\bibitem{NumericalRecipes}W.~H.\ Press, S.~A.\ Teukolsky, W.~T.\ Vetterling, and B.~P.\ Flannery, {\em Numerical Recipes} (Cambridge University Press, 2007), 3rd ed.
\bibitem{Dieckmann:1998}K.\ Dieckmann, R.~J.~C.\ Spreeuw, M.\ Weidem\"{u}ller, and J.~T.~M.\ Walraven, \pra {\bf 58}, 3891 (1998).
\bibitem{Chaudhuri:2006}S.\ Chaudhuri, S.\ Roy, and C.~S.\ Unnikrishnan, \pra {\bf 023406} (2006).
\bibitem{Cren:2002rt}P.\ Cren, C.~F.\ Roos, A.\ Aclan, J.\ Dalibard, and D.\ Gu\'{e}ry-Odelin, Eur.\ Phys.\ J.\ D {\bf 20}, 107 (2002).
\bibitem{Lu:1996}Z.~T.\ Lu, K.~L.\ Corwin, M.~J.\ Renn, M.~H.\ Anderson, E.~A.\ Cornell, and C.~E.\ Wieman, \prl {\bf 77}, 3331 (1996).
\bibitem{Slowe:2005}C.\ Slowe, L.\ Vernac, and L.~V.\ Hau, Rev.\ Sci.\ Instrum.\ \textbf{76}, 103101 (2005).
\bibitem{Streed:2006}E.~W.\ Streed, A.~P.\ Chikkatur, T.~L.\ Gustavson, M.\ Boyd, Y.\ Torii, D.\ Schneble, G.~K.\ Campbell, D.~E.\ Pritchard, and W.\ Ketterle, Rev.\ Sci.\ Instrum.\ {\bf 77}, 023106 (2006).
\bibitem{Muller:2009}T.\ M\"{u}ller, M.\ Gilowski, M.\ Zaiser, P.\ Berg, C.\ Schubert, T.\ Wendrich, W.\ Ertmer, and E.~M.\ Rasel, Eur.\ Phys.\ J.\ D {\bf 53}, 273 (2009).
\bibitem{Proukakis:2003}N.~P.\ Proukakis, Laser Phys. {\bf 13}, 527 (2003).
\bibitem{Zaremba:1999}E.\ Zaremba, T.\ Nikuni, and A.\ Griffin, J.\ Low Temp.\ Phys.\ {\bf 116}, 277 (1999).
\bibitem{Blakie:2008a}P.~B.\ Blakie, A.~S.\ Bradley, M.~J.\ Davis, R.~J.\ Ballagh, and C.~W.\ Gardiner, Advances in Physics \textbf{57}, 363 (2008).
\bibitem{Burt:1997fk}E.~A.\ Burt, R.~W.\ Ghrist, C.~J.\ Myatt, M.~J.\ Holland, E.~A.\ Cornell, and C.~E.\ Wieman, \prl {\bf 79}, 337 (1997).
\bibitem{Soding:1999}J.\ S\"{o}ding, D.\ Gu\'{e}ry-Odelin, P.\ Desbiolles, F.\ Chevy, H.\ Inamori, and J.\ Dalibard, Applied Physics B: Lasers and Optics {\bf 69}, 257 (1999).
\bibitem{Jack:2002}M.~W.\ Jack, \prl {\bf 89}, 140402 (2002).
\bibitem{Wick:1950}G.~C.\ Wick, Phys.\ Rev.\ \textbf{80}, 268 (1950).


\end{thebibliography}
\end{document}